\documentclass[twocolumn]{aastex631}

\usepackage{xspace}

\newcommand{\nicer}{\textit{NICER}\xspace}

\newcommand{\cxo}{{\it Chandra}\xspace}

\newcommand{\xmm}{{\it XMM-Newton}\xspace}

\newcommand{\swift}{{\it Swift}\xspace}

\newcommand{\nustar}{{\it NuSTAR}\xspace}
\newcommand{\nus}{{\it NuSTAR}\xspace}

\newcommand{\ergs}[1]{$\times 10^{#1}$ erg s$^{-1}$}

\newcommand{\ngc}{SN\,2010da/\allowbreak NGC\,300\allowbreak\,ULX-1\xspace}
\newcommand{\snda}{SN\,2010da\xspace}
\newcommand{\ulx}{NGC\,300\,ULX-1\xspace}

\newcommand{\Rev}[1]{{ #1}}

\graphicspath{{./}{plots/}}

\begin{document}

\title{Detection of the optical counterpart of the transient ULX NGC300 ULX-1:  a nascent black hole -- neutron star binary?}

\author[0000-0002-1115-6559]{Andr\'e-Nicolas Chen\'e}
\affiliation{NSF NOIRLab,  670 N. A'ohoku Place, Hilo, HI, 96720, USA}

\author[0000-0003-3902-3915]{Georgios Vasilopoulos}
\affiliation{Department of Physics, National and Kapodistrian University of Athens, University Campus Zografos, GR 15784, Athens, Greece}
\affiliation{Institute of Accelerating Systems \& Applications, University Campus Zografos, Athens, Greece}

\author[0000-0003-0708-4414]{Lidia M. Oskinova}
\affiliation{Institut f\"ur Physik und Astronomie, Universit\"at Potsdam, Karl-Liebknecht-Str. 24/25, 14476 Potsdam, Germany}

\author[0000-0002-9144-7726]{Clara Mart\'inez-V\'azquez}
\affiliation{international Gemini Observatory/NSF NOIRLab,  670 N. A'ohoku Place, Hilo, HI, 96720, USA}



\begin{abstract}

The end points of massive star evolution are poorly known, especially those in interacting binary systems containing compact objects, such as neutron stars or black holes. Such systems are bright in X-rays, and the most luminous among them are called ultra-luminous X-ray sources (ULXs).
In this paper, we address the enigmatic \ulx. It's X-ray activity started in 2010 with the supernova impostor-like event \snda. In the following few years the ULX  was powered by persistent super-Eddington accretion but then it dimmed in X-rays. We present the most recent X-ray and optical observations. The {\em Chandra} and {\em Swift} telescopes confirm that \ngc is not accreting at super-Eddington level anymore. We attribute this switch in accretion regime to the donor star variability and its fast evolution. In order to gain a better understanding of  the donor star's nature, we consider its optical light curve on a decade-long time scale and show that the optical counterpart of \ngc dimmed significantly over recent years. The most recent detection in optical by the Gemini telescope reveals that the source is now $>2.5$\,mag fainter in the $r'$ band compared to the epoch when it was spectroscopically classified as a red supergiant. We  discuss the nature of the  abrupt changes in the donor star properties, and consider among other possibilities the silent collapse of the donor star into a black hole.

\end{abstract}

\keywords{}

%% and \citet commands to identify citations. The citations are

\section{Introduction} \label{sec:intro}

High-mass X-ray binaries (HMXBs) with X-ray luminosities above $\sim 10^{36}$\,erg\,s$^{-1}$ (in the $0.2-60$\,keV range) are powered by mass transfer from a massive donor star to a neutron star (NS) or black hole (BH). HMXBs can be persistent, but more commonly, they are transient X-ray sources.

%%%%%%%%%%%%%%%%%%%%%%%%%%%%%%% Fig 1 %%%%%%%%%%%%%%%%%%%%%%%%%%%%%%%%%
\begin{figure*}
    \centering
    \includegraphics[width=0.49\linewidth]{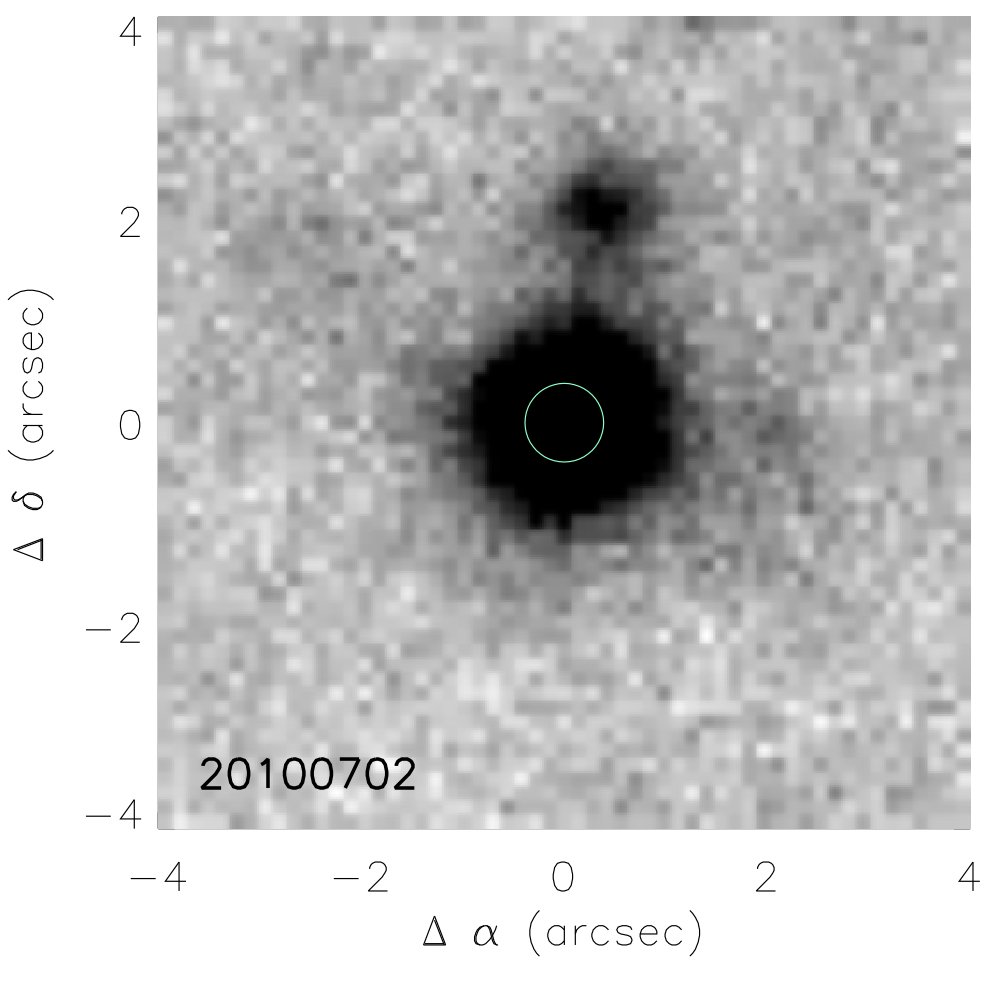}
    \includegraphics[width=0.49\linewidth]{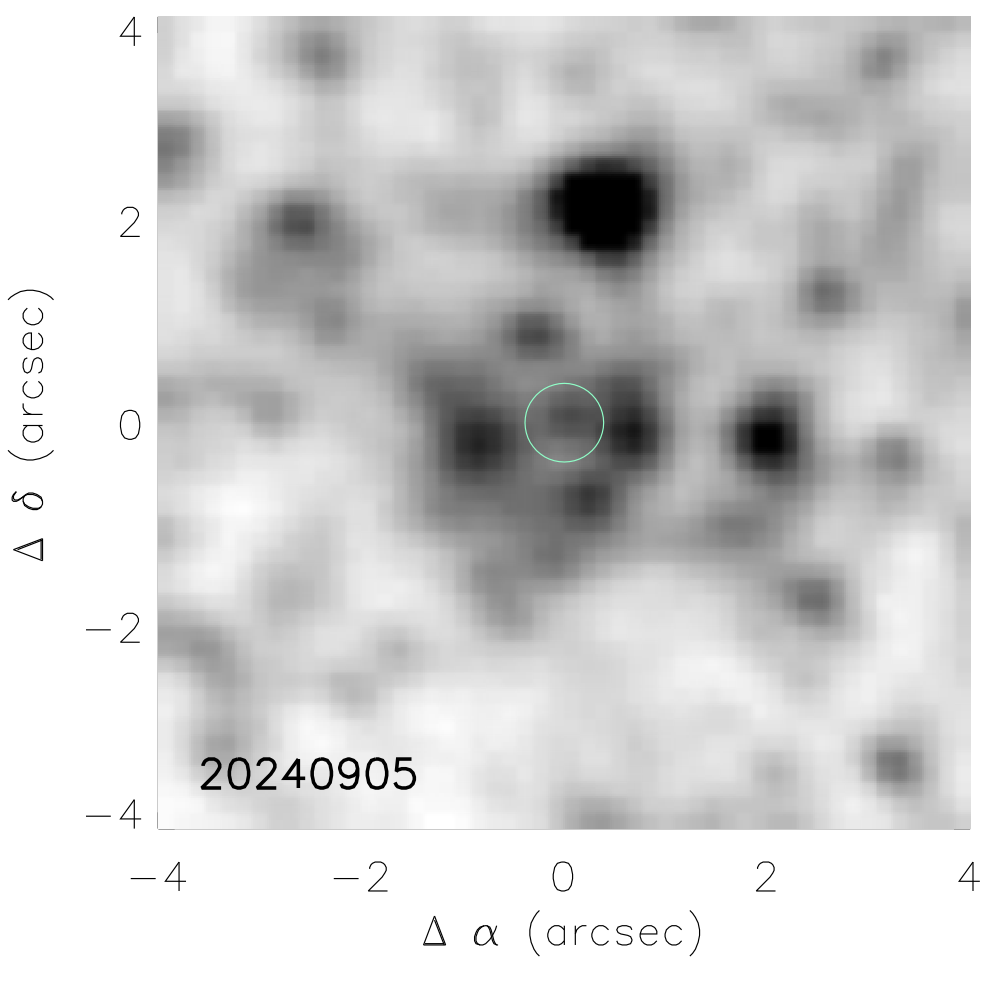}
    \caption{Cut-out from the $r'$-band GMOS-South images of \ngc taken in 2010 (left) and 2024 (right). The light blue circle marks the position of the source.}
    \label{fig:fov}
\end{figure*}
%%%%%%%%%%%%%%%%%%%%%%%%%%%%%%%%%%%%%%%%%%%%%%%%%%%%%%%%%%%%%%%%%%%%%%%

HMXBs with $L_X\gtrsim 10^{39} \, \rm erg \, s^{-1}$ are classified as ultra-luminous X-ray sources (ULXs) \citep[see review][]{King2023}. The discovery of ULX pulsars \citep[ULXPs, e.g][]{2014Natur.514..202B} showed that, at least some ULXs are fueled by super-Eddington accretion onto a NS. However neither the duty cycle of ULXPs nor their evolution and origin are fully understood. Most ULXs/ULXPs exhibit modulation in their X-ray light curves, which is attributed to variability in the outflow that influences the collimation of X-rays along the line of sight \citep[e.g.][]{2021A&A...654A..10G,2021ApJ...909...50V}. Hence, sources may transition between ULX and normal accretion phases, as seen in the recently discovered transient ULX with an OBe-type donor in the Galaxy \citep{Reig2020}. 

Changes in the accretion outflow a onto the NS could be caused by  the variability  of the donor star. Indeed, most evolved massive stars are variable by nature. 
A dramatic historic example is the so-called Great Eruption of the Galactic luminous blue variable (LBV) star $\eta$\,Carinae. The star increased optical brightness by $\sim 8$\,mag,  ejected a large amount of matter, and subsequently returned to its  pre-eruption brightness. In general, the LBV eruptions are energetic enough to be misclassified as \Rev{supernovae (SNe)}. This led to the introduction of a new type of transients, the so-called SN impostors \citep{Mauerhan2013}.  

One enigmatic SN impostor, \snda, is associated with the transient ULX named \ulx (from now on \ngc). 
%At the time of its discovery in 2010, 
The intense multiwavelengths observations of \ngc over the last 15 years have revealed its remarkable transformations.
The system which was once classified as a persistent ULX where a NS accretes matter from  a red superginant (RSG) donor \citep{Heida2019} is now quiescent.  
Here, we present the latest X-ray and optical observations of \ngc, along with the contemporaneous detection of its optical counterpart.

This paper is organized as follows. In Section\,\ref{sec:obs} we present new $r'$-band and X-ray monitoring observations as well as the details about constructing long term light-curves in $r'$-band and X-rays. Section\,\ref{sec:chrono} outlines the chronology of events and summarizes observations in X-ray and optical. Sections\,\ref{sec:discussion} and \ref{sec:conclusion} present our discussion and conclusions.

\section{Observations} \label{sec:obs}

\subsection{Optical imaging}
\label{sec:opt}

In September 2024, we obtained imaging observations of \ngc using the Gemini South telescope under the program GS-2024B-FT-104. A total exposure time of 2400s was achieved with the GMOS-South instrument \citep{2004PASP..116..425H} in imaging mode, using the $r'$ Sloan filter and the new Hamamatsu CCDs \citep{2016SPIE.9908E..2SG}. The data were processed and combined using the DRAGONS software package \citep{2023RNAAS...7..214L}. The section of the image around \ngc is shown in the right panel of Figure\,\ref{fig:fov}.

Since 2010, the sky field around \ngc has been observed multiple times with different telescopes and instruments, providing valuable archival data for our analysis. Observations in the $r'$ filter with GMOS-South on the Gemini Observatory are available through the Gemini Observatory Archive. These include a 60-second exposure from July 2010 (program GS-2010-Q-19; left panel of Figure\,\ref{fig:fov}) and additional acquisition images taken incidentally during observations of nearby sources. While these acquisition images often have increased noise, shorter exposure times, and suboptimal calibration due to fast readout modes and unstable optics, they remain useful for constructing a historical light curve.

Similarly, an acquisition image obtained with the FORS2 instrument on the Very Large Telescope-U1 (program 105.20HJ) in September 2021, using the $GG435$ filter, contributes to the photometric dataset. Despite the varied origins and quality of these archival images, they collectively provide sufficient accuracy to support our current analysis.

\begin{figure*}
    \centering
    \includegraphics[width=0.99\linewidth,trim={0 11 0 5 }, clip]{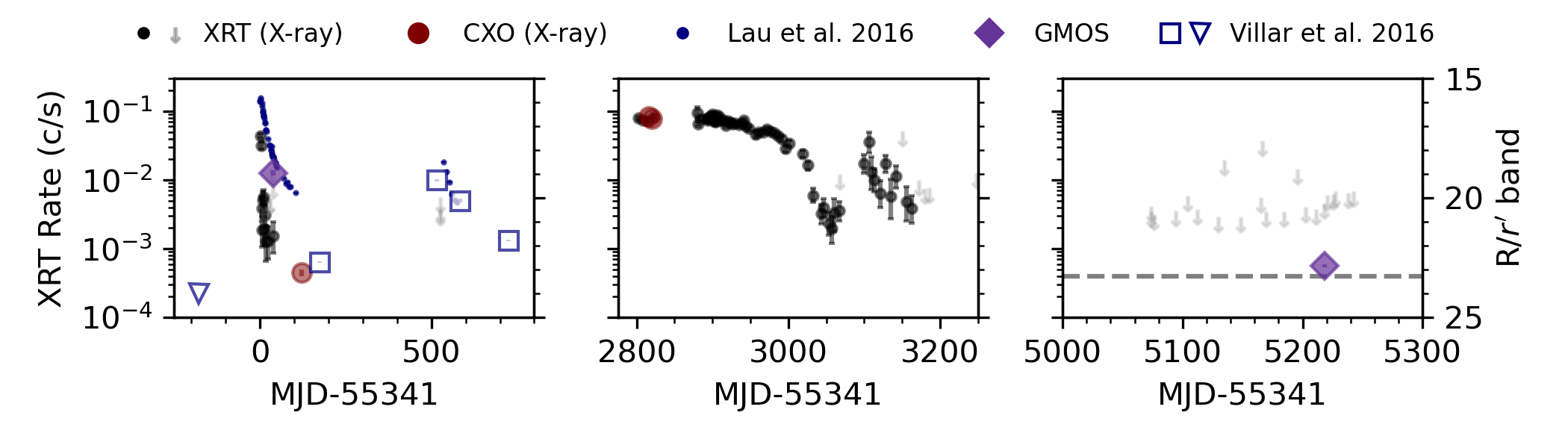}
    \caption{
    %\gv{2 plot options, see X-scale around 2010-11, and Villar data in 2nd.} 
    Optical and X-ray light curve zoomed in over 3 epochs. \emph{Left panel:} the \snda event. \emph{Middle panel:} the 2018 campaign that followed the discovery of pulsations and characterization of the system as a ULXP (middle panel). \emph{Right panel:} \swift/XRT pointing over 2024. \swift/XRT 3$\sigma$ upper limits are marked with downward gray arrow (values correspond to mid point of arrows). Horizontal hashed line marks upper limit from stacked observations of the 2024 monitoring data. }
    \label{fig:lcz}
\end{figure*}

The photometric reduction was performed using {\sc daophot/allstar} \citep{1987PASP...99..191S, 1994PASP..106..250S} following similar prescriptions as those described in \citealt{2021AJ....161..120M}. The flux calibration was made using the DECam Local Volume Exploration Survey Data Release 2
 \citep[DELVE DR2,][]{2022ApJS..261...38D}\footnote{DELVE DR2 data was downloaded from the Astro Data Lab which is part of the Community Science and Data Center (CSDC) at NSF NOIRLab, the national center for ground-based nighttime astronomy in the United States operated by the Association of Universities for Research in Astronomy (AURA) under cooperative agreement with the U.S. National Science Foundation.}. Unfortunately, the FORS2 acquisition image could not provide better than an upper limit due to low signal. The resulting light curve is presented in Figure\,\ref{fig:lcz} and \ref{fig:lc}, and the extracted values are available in Table\,\ref{tab:rPhot}.

%%%%%%%%%%%%%%%%%%%%%%%%%%%%%  Figure 2 %%%%%%%%%%%%%%%%%%%%%%%%%%%%%%%%
%%%%%%%%%%%%%%%%%%%%%%%%%%%%%%%%%%%%%%%%%%%%%%%%%%%%%%%%%%%%%%%%%%%%%%%

\subsection{X-ray observations}
\label{sec:xray}

\begin{figure*}
    \centering
    \includegraphics[width=0.9\linewidth,trim={0 10 0 10 }, clip]{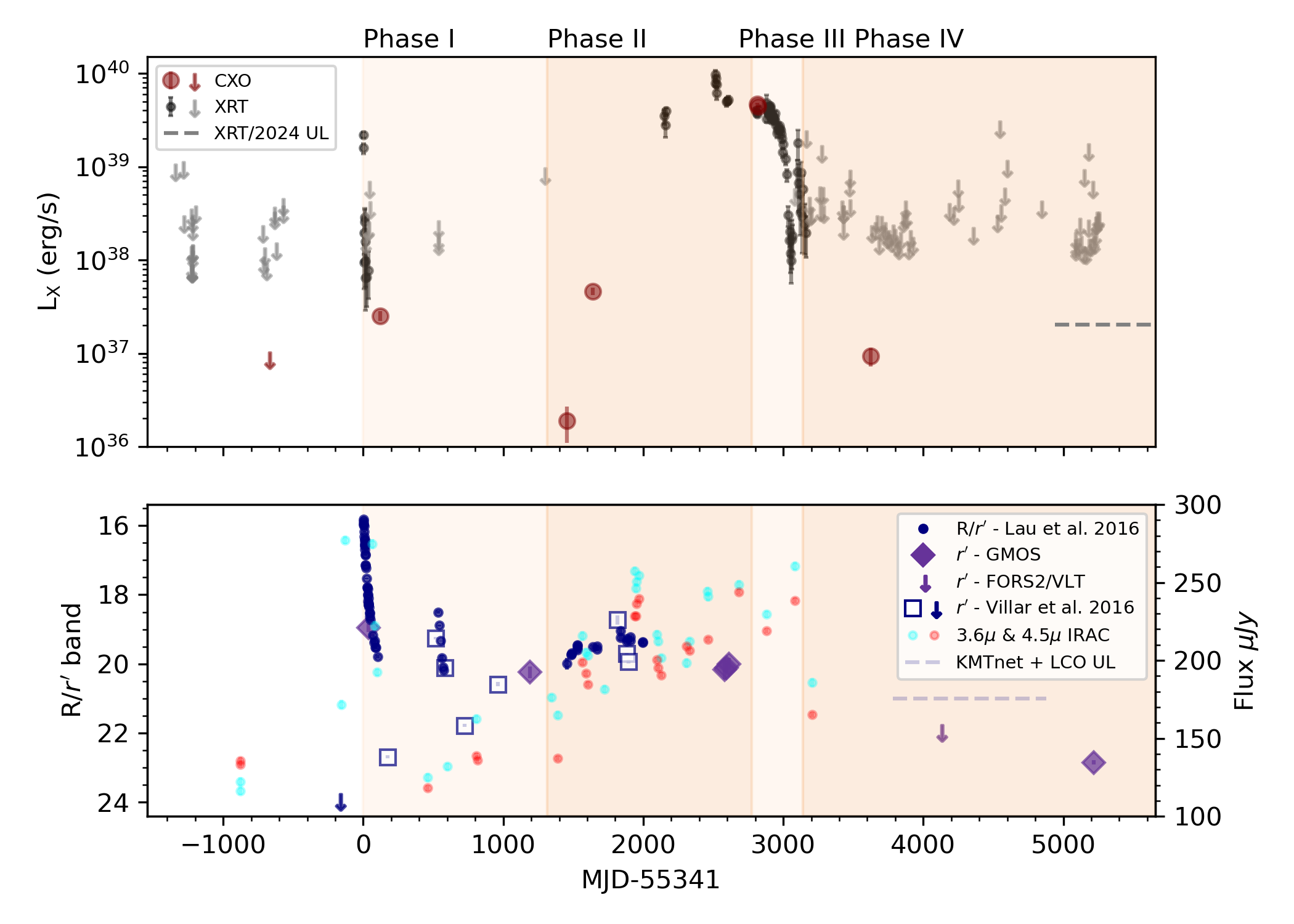}
    %\vspace{-1cm}
    \caption{Long term X-ray light ({\em Upper panel}) and optical ({\em Lower panel}) light curve of \ngc. Characteristic phases following the 2010 impostor event are marked with shaded regions based on X-ray behavior. The $L_{\rm X}$ is inferred from the count rates assuming constant bolometric correction, i.e. not accounting for changes in absorption.
    Following the impostor event the X-ray flux drops, while in phase II the flux increases and the source breaks the ULX limit before the flux starts to gradually drop within 2018 (phase III). Within phases II and III there are strong indications that changes in intensity are mainly due to absorption (see text). Within phase IV the source is only detected with deep X-ray observations, and only upper limits were obtained by XRT snapshots or stacked observations. Optical magnitudes are taken from the literature (see text), apart from GMOS which we analyzed. The dashed line shows an upper limit determined by the MKTnet and LCO archival data.}
    \label{fig:lc}
\end{figure*}

Following the discovery of X-ray pulsations in 2018, \ngc was extensively monitored by \swift/XRT (0.2--10\,keV) and \nicer\ (0.2--12\,keV) X-ray telescopes \citep[see][]{2018A&A...620L..12V,2019ApJ...879..130R} (see Figure\,\ref{fig:lcz}). Around August 2018 the X-ray flux dropped, and after some fluctuation the system entered a low flux state \citep{2019MNRAS.488.5225V}. Since that time we have continued monitoring \ngc with \swift/XRT aiming at detecting a possible rebrightening. We note that \nicer observations are only useful during the brightest stage of the system and even then are contaminated by nearby system NGC 300 X-1 \citep{2022ApJ...940..138N}. The field of sky around \ngc has also been observed with \nustar (3--79\,keV), \cxo (0.2--12\,keV) and \xmm (0.2--12\,keV) X-ray telescopes. However, among these observatories only \cxo allowed meaningful detections at low flux, significantly below the lower limits of \swift/XRT individual snapshots. Hence, to gain insights on the behavior of \ngc over time scale of a decade, we construct the X-ray light curve using \swift/XRT and \cxo data (Figures.\,\ref{fig:lcz} and \ref{fig:lc}).

\begin{table}[]
    \centering
    \caption{X-ray count rates in the 0.5-8.0\,keV range and $r'$ photometry of \ngc obtained from from CXO and GMOS-South, respectively.}
    \begin{tabular}{lr@{$\pm$}l|lr@{$\pm$}l}
    \hline\hline
    \noalign{\smallskip}
    \multicolumn{3}{c}{X-Ray - (0.5-8.0 keV)} &\multicolumn{3}{c}{Optical}\\
    \multicolumn{1}{c}{Date} & \multicolumn{2}{c}{($10^{-3}$ c/s)} & \multicolumn{1}{c}{Date} & \multicolumn{2}{c}{$r'$ (mag)}\\
    \hline
    \noalign{\smallskip}
    2008 Jul 08 & \multicolumn{2}{c}{$<$0.4}  & 2010 Jul 02  &  18.96 & 0.11\\
    2010 Sep 24 & 1.20 & 0.14 & 2013 Aug 26  &  20.24 & 0.17\\
    2014 May 16 & 0.09 & 0.04 & 2017 Jun 19  &  20.16 & 0.29\\
    2014 Nov 17 & 2.21 & 0.19 & 2017 Jul 02  &  20.11 & 0.13\\
    2018 Feb 08 & 225  &   5  & 2017 Jul 18  &  20.00 & 0.17\\
    2018 Feb 11 & 210  &   5  & 2021 Sep 01  & \multicolumn{2}{c}{$<$22$^\dagger$} \\
    2020 Apr 26 & 0.45 & 0.10 & 2024 Sep 05  &  22.85 & 0.06\\
    \noalign{\smallskip}
    \hline\hline
    \end{tabular}
    \label{tab:rPhot}
    \begin{minipage}{13cm}
        \small $^\dagger$: upper limit evaluated from the photometry of the FORS2\\ image, converted from the $GG$435 to the $r'$ filter.
    \end{minipage}
\end{table}

Archival data were analyzed using {\tt CIAO v4.16} via {\tt conda} with {\tt CALDB v4.11.0}. We constructed clean images in the 0.5-8.0 keV band and performed source detection via {\tt wavdetect}. 
The latest \cxo pointing was performed on April 26 2020 (obsid: 22375, PI: B. Binder), where the source was detected with a rate of 4.4(1.0)$\times$10$^{-4}$ c/s.
The \swift/XRT data of sky filed around \ngc\ were analyzed and a long term X-ray light curve was produced using standard procedures \citep{2007A&A...469..379E,2009MNRAS.397.1177E}. 
The \swift/XRT monitoring of \ngc over 2019-2024 (PI: Vasilopoulos, G.) has a typical cadence of one or two months. During this period, the source is not detected and upper limits could be established. In 2024, a few quite deep observations of the NGC\,300 galaxy included the sky field around \ngc. These 
data were stacked in order to obtain an image with a combined exposure time 
65\,ks. We run source detection on the stacked image and obtained only upper limits for the location of the \ngc\ with a rate of 0.0004 counts\,s$^{-1}$, which is a factor of 10-100 lower compared to the upper limits obtained by the individual XRT snapshots. In order to compute the light-curve we follow the procedure described in \citet{2019MNRAS.488.5225V} and use liner scaling relations to determine the upper limit on $L_{\rm X}$ (0.3-30 keV) 
assuming the same spectral model as in \citet{2018MNRAS.476L..45C}. 
The conversion of the count rate, presented in Table\,\ref{tab:rPhot}, to flux was done using the factors
2.1\ergs{40}\,/cnt\,s$^{-1}$ and 5.6\ergs{40}\,/cnt\,s$^{-1}$ for \cxo and XRT, respectively.

To better show the short term evolution of the X-ray light curve we plot individual epochs of interest in Figure\,\ref{fig:lcz}, while the complete light-curve is presented in Figure\,\ref{fig:lc}. In all plots, the time zero is scaled such zero is at the time of the SN\,2010da outburst.
% In previous section we have already discussed the different phases in the evolution of \ngc as characterized mainly by X-ray.

%Here for the first time we have obtained an updated light-curve that characterizes its evolution after 2018 and the last time where pulsations were detected. 
We present here an updated light curve characterizing the evolution of the X-ray luminosity of \ngc following 2018. This marks the period after the last detection of X-ray pulsations from the source
% In Fig. \ref{fig:lc} we show that during what now seems like a new evolutionary phase, the system is steadily below the Eddington limit (albeit constrained by upper limits), and the only robust detection comes by \cxo. The 2020 \cxo rate translates to a flux of \oergs{37}, while XRT upper limit from stacked data is a factor of 2 higher than \cxo detection. 
% Nevertheless, these limits were obtained assuming no strong X-ray absorption, and given that no X-ray spectra can be obtained over this 6 year period we cannot exclude the possibility that the flux of \ngc is significantly absorbed by prolonged obscuration and/or absorption.

\section{From Outburst to Aftermath} \label{sec:chrono}

To place the new observations in the context of the longer term evolution, in this section we briefly recap the chronology of events starting shortly before the SN impostor SN\,2010da event \citep[see in depth reviews in][]{Villar2016,Binder2020}. For clarity, 
we divide these in four phases as illustrated in Figure\,{\ref{fig:lc}}.

\subsection{Phase I. 2010 -- 2014}

%The galaxy NGC\,300 is frequently observed across the electromagnetic spectrum.
Before 2010, no source had been detected at the position of \ngc at wavelengths from X-rays \citep[upper limit of $<10^{37}$\,erg\,s$^{-1}$, ][]{Binder2011} to optical \citep[conservative limits of $24$\,AB mag,][]{Berger2010}. However, in 2007, the {\em Spitzer}  telescope detected a mid-IR source which brightened by $\sim 0.5$\,mag within six months before the 2010 event \citep{Lau2016}.

On May 25, 2010, an optical transient, appeared at $r'\approx 16$\,mag. Although initially classified as a SN, getting the name \snda, its energetics were much lower than typical for such events. Inspection of the {\em Spitzer} data shows that the mid-IR flux had raised just before the optical outburst and sharply dropped soon afterward. The X-ray luminosity raised to 5\ergs{38} at the time of the outburst, but has declined by a factor of 25 within months after \citep{Binder2011}.
%The initial interpretation of the transient as an LBV-like eruption obscured by dust was questioned by various authors. \citet{Lau2016} suggested that the source is a B[e]-HMXB binary, and discussed alternative scenarios such as a stellar merger or a low-luminosity SN.
 
% \citet{2016MNRAS.457.1636B} suggested that we have directly witnessed the formation of a HMXB.

\subsection{Phase II. 2014\,--\,2018}
\label{sec:ulx}

During 2014 the system remained X-ray active, though at lower levels. But by 2016, \nus and \xmm observations established that the source became an ULX with the X-ray luminosity $\approx 5$\ergs{39}. X-ray pulsations were detected and a spin-up rate of $-5.56\times10^{-7}$\,s\,s$^{-1}$ was established. The NS spin rapidly evolved from $\sim$126\,s in 2014 to $\sim$31.5\,s in 2016 and to $\sim 16.5$\,s in November 2018. Temporal analysis did not reveal any evidence for orbital modulations, implying either an unlikely pole-on orbital configuration for the system or a large ($>1$\,yr) orbital period \citep{2018MNRAS.476L..45C,2018A&A...620L..12V,2019MNRAS.488.5225V}.

X-ray spectral analysis found signs of ultra-fast outflows and a possible cyclotron resonance feature, suggesting a strongly magnetized NS \citep[$\gtrsim10^{12}$\,G;][]{2018ApJ...857L...3W}. Comparison of X-ray spectra from 2010 and 2016 showed similar continua but much lower obscuration in 2016 compared to 2010, shortly after the \snda impostor event \citep{2018MNRAS.476L..45C}. 
%Modeling indicated that the 2010 outburst destroyed much of the surrounding dust, with newly formed dust appearing closer to the system \citep{Lau2016}.

%{\it Hubble} observations identified an optical counterpart of the ULX, showing that the donor star is likely a blue supergiant \citep{Villar2016}.% REF??
%, possibly associated with young stars aged $\le$ 5 Myr. 

%This finding shows that a spin equilibrium period can be reached within a matter of a couple years for extreme accretion, thus showing that an event like this could not have happened again for \ngc, since had it occurred, the system could not had been rotating so far away from its equilibrium period to begin with.

\subsection{Phase III. 2018}

By 2018, the optical counterpart of \ngc\ had $r'\le20.1$\,mag.  \citet{Heida2019} obtained its optical and near-IR spectra and classified the source as a RSG.  
%Based on single star evolutionary tracks, the relatively low luminosity ($\log{L_\ast/L_\odot}\approx 4.25$) implies a donor star mass of $10M_\odot < M_{\rm donor} < 15M_\odot$. 
The binary orbital period longer than a year is consistent with the the RSG radius \citep[][]{2019ApJ...879..130R}.

Later in 2018, the X-ray flux starts to strongly decline
(by a factor of 50 within a year), while  the NS spin-up remained constant. This implied that accretion onto the NS has continued but possibly became strongly obscured  \citep{2019MNRAS.488.5225V}. Mid-IR emission also sharply declined, possibly explained by the heating of the dust by X-rays.
 
\subsection{Phase IV. 2019\,--\,now.}

Starting from 2019, the system has further dimmed in X-ray, IR, and optical. %Our new observations confirm that \ngc entered a quiescence phase that is still ongoing.
Continuing monitoring with {\em Swift}/XRT returns only upper limits on  X-ray luminosity.
%does not exceed $10^{38}$\,erg\,s$^{-1}$. 
The last X-ray detection in 2020 (sect.\,\ref{sec:xray}) measured  luminosity of $L_{\rm X} \approx 10^{37}$\,erg\,s$^{-1}$. This suggests that accretion onto the NS was still ongoing at that time, though no longer in a supercritical regime.
%The limited number of detected counts prevents extracting a spectrum or a pulsation light curve. 

Beside the FORS2 archival image and our new $r'$-band GMOS-South imaging (sect.\,\ref{sec:opt}), there are no convincing detections neither in optical or IR since 2019. Archival data from the Korean Microlensing Telescope network (KMTnet) and from the Las Cumbres Observatory (LCO) obtained between 2020 and 2021 show some signal that is compatible to an unresolved detection of the  group of sources in the vicinity of \ngc (as seen in Figure\,\ref{fig:fov}). We therefore assume a conservative upper limit of $\sim$21\,mag during that time. The upper limit is estimated as the minimum brightness \ngc would need to reach in order to result into an increased total brightness of the combined signal coming from all the sources within the point spread function (PSF) of the KMTnet and LCO archival data.

\section{Discussion} \label{sec:discussion}

%In this section, we are examining the new detection in the $r'$-band in relation to the most recent X-ray detections and prior observations of NGC300\,ULX-1. This analysis helps to establish if the system's current state reflects its pre-2010 conditions, or if it was transformed by this nearly one decade long high activity period. Without comprehensive color data or a spectrum, we are limited in our ability to answer that question.

%In 2024, the brightness of the optical counterpart of \ngc is only marginally brighter compared to the pre-outburst, while its X-ray flux has dropped below detection limit of monitoring {\em Swift}/XRT observations. 
%Our new data show that the system remains in a quiescent state since 2019. %What these new results tell about the nature of \ngc and what happened over the last 15 years? 
The fast evolution of \ngc\ during last decade is likely governed by mechanisms acting on short time scales. 
%This contrasts sharply with  persistent ULXs, which are powered by stable Roche lobe overflow . 
The new data reported in this paper help to elucidate the nature of these processes. 
%further develop on the recent evolution of \ngc.

\subsection{NS spin period and the 2010 event}

X-ray pulsations were discovered in 2014  with long period of 126\,s. This suggests that the NS was formed long before the \snda event 
because young NS typically have spin periods on the order of only 0.1\,s or shorter \citep{2022MNRAS.514.4606I}.
%
%because young NS typically have spin periods of just a few %milliseconds \citep{2022MNRAS.514.4606I}. 
%
An alternative scenario where the primary star did collapse into a NS during the \snda event but was spun down rapidly
seems very unlikely. While spin reversal due to steady accretion \citep{2018A&A...620L..12V} can help to explain slow  rotation,  this scenario challenges standard NS formation models and angular momentum conservation in binaries. 
%We therefore conclude the 2010 event was related to the secondary star, as suggested by most previous studies of the system.

The slow NS spin also implies that the super-Eddington accretion episode that powered the ULX between 2010-2018 took place for the first time in the history of \ngc.  Indeed, if it had happened repeatedly, the NS would have already spun-up to shorter periods before 2014.  

\subsection{Donor star variability}

The evolution of light curves across wavelengths from X-rays to mid-IR is linked to the secondary star's activity. Here, we present three possible scenarios.

\subsubsection{RSG variability scenario}

Following \citet{Heida2019} and assuming the donor star is an RSG, the increase in X-ray luminosity observed in 2014 can be explained by a phase of excessive super-Eddington accretion. This may have been triggered by episodic mass loss or a sudden increase in the donor star's radius. Associated dust formation and destruction, along with resulting changes in extinction, further complicate the observed picture.

Mass-loss episodes are frequently observed in RSGs. For example, the well-studied R [W60] B90, RW Cep, $\mu$ Cep, and Betelgeuse exhibit episodic mass loss with typical timescales of 200-400 days \citep[e.g.,][]{2024A&A...690A..99M}.

\subsubsection{Radius expansion}

Sudden radius expansion of the donor star is also a possibility. A relevant scenario was proposed by \citet{Gilkis2019}, who considered a case where the optical star in a binary system containing an NS rapidly expands due to core instability near the end of its nuclear evolution. As the NS plunges into the stellar envelope, it begins accreting material at a super-Eddington rate, initiating a transient outburst. Once jets are launched, the donor star may either survive or be disrupted.

A similar scenario could apply to \ngc if the NS in this system is on a highly elliptical orbit due to a kick. After the donor star expands, the next periastron passage may lead to the NS grazing the donor star's envelope, triggering the SN impostor event in 2010. The subsequent periastron passage in 2016 could manifest as an ULX phase, eventually leading to the disruption of the donor star.

However, it seems unlikely that the NS and the donor star, hypothetically a RSG, merged around 2020 to form a Thorne-\.Zytkow object \citep[T\.ZO,][]{1975ApJ...199L..19T}, as this does not readily explain the optical dimming. Furthermore, the luminosity of the RSG appears too low for a T\.ZO \citep{Farmer2023}.

\subsubsection{Direct collapse of the donor star}

We speculate that the 2010 SN impostor-like event was indicative of the donor star becoming unstable, a phenomenon often observed among core-collapse SN progenitors \citep[e.g.,][]{Qin2024}. This may have led to a brief period of super-Eddington accretion onto the NS, during which the system was observed as an ULX. Accretion ceased when the donor star collapsed without producing an associated SN \citep{Sukhbold2020}. Although the "failed SN" resulted in some mass ejection, a fraction of the material likely fell back onto the proto-NS, leading to the formation of a BH. The X-ray source observed by {\em Chandra} in 2020, along with the optical counterpart, was powered by residual accretion.

In this scenario, the current optical source is analogous to failed SN remnants detected in the M31 and NGC 6946 galaxies \citep{2024arXiv240403719B,De2024}. Notably, NGC 6946-BH1 experienced a luminous optical outburst in 2009 \citep{2017MNRAS.468.4968A}, followed by the formation of an expanding dusty envelope over subsequent years \citep{2021MNRAS.508.1156B}. The galactic V404 Cyg triple system is another candidate where evidence suggests BH formation with minimal kick velocity, possibly resulting from a failed supernova \citep{2024arXiv240403719B}.

If confirmed through future optical and X-ray monitoring, \ngc would represent the first known example of a failed supernova leading to the formation of a BH-NS binary.

\section{Summary and conclusion} \label{sec:conclusion}

%\ngc is a massive binary system observed at a rare and brief evolutionary stage. The primary star had already collapsed into a NS by the time of the \snda event. Between 2018 and 2020, the NS experienced a period of supercritical accretion, resulting in significant spin-up. During this phase, \ngc appeared as an ULX, but its X-ray brightness has since declined by at least two orders of magnitude.

%The secondary star, serving as the optical donor, appears to be nearing final core-collapse. Initially classified as an sgB[e] star and later as a red supergiant (RSG), its optical brightness has steadily declined over the past five years. Its current spectral type remains unknown.

%In this paper, 
We report the latest X-ray and optical observations of \ngc. The system was last detected in X-rays by the {\em Chandra} telescope in 2020, with a luminosity of $L_{\rm X} \sim 10^{37}$\,erg\,s$^{-1}$, indicating ongoing accretion onto the NS. However, the low flux prevents meaningful spectral or timing analysis of this data. The on-going monitoring with the {\em Swift} telescope shows that super-Eddington accretion has \Rev{ceased}. 
%, which are insufficient to determine if accretion continues or has ceased.

In 2024, we successfully detected the optical counterpart of \ngc using the Gemini telescope. The source is approximately 2.5 magnitudes fainter in the $r'$-band compared to the last optical detection prior to 2019. %New multi-wavelength observations are required to further investigate the nature of the optical counterpart.

We propose two possible \Rev{explanations for} the observed phenomenology of \ngc. One possibility is the interaction of the NS with the supergiant donor\Rev{, which may have experienced a change in its mass loss or radius. Alternatively, we speculate that} within the past five years, the core of the secondary star may have collapsed without producing  a SN, suggesting that \ngc has evolved into a relativistic NS$+$BH binary. \Rev{

These scenarios will be tested with future observations. If an X-ray binary with $L_{\rm X}>10^{35}$\,erg\,s$^{-1}$ (as common for NSs accreting from stellar winds) and a stellar counterpart are detected, it would rule out the NS+BH scenario. If the donor star has survived, it should appear either as a hot (bluish) stripped star undergoing a relaxation process or as a cool, RSG-type object. Conversely, if the spectral energy distribution of the source is not stellar and any persistent accretion is moderate, likely from a residual circumstellar nebula, it would be in favor of the NS+BH scenario.}

%% Also note that the akcnowlodgment environment does not support long amounts of text. If you have a lot of people and institutions to acknowledge, do not use this command. Instead, create a new \section{Acknowledgments}.
\begin{acknowledgments}
We thank Avishai Gilkis for the discussion on failed SNe and NS formation. 
GV acknowledges support from the Hellenic Foundation for Research and Innovation (H.F.R.I.) through the project ASTRAPE (Project ID 7802).

The scientific results reported in this article are based on observations made by the Chandra X-ray Observatory.

Based on observations obtained at the international Gemini Observatory, a program of NSF NOIRLab, which is managed by the Association of Universities for Research in Astronomy (AURA) under a cooperative agreement with the U.S. National Science Foundation on behalf of the Gemini Observatory partnership: the U.S. National Science Foundation (United States), National Research Council (Canada), Agencia Nacional de Investigaci\'{o}n y Desarrollo (Chile), Ministerio de Ciencia, Tecnolog\'{i}a e Innovaci\'{o}n (Argentina), Minist\'{e}rio da Ci\^{e}ncia, Tecnologia, Inova\c{c}\~{o}es e Comunica\c{c}\~{o}es (Brazil), and Korea Astronomy and Space Science Institute (Republic of Korea). Archival data were acquired through the Gemini Observatory Archive at NSF NOIRLab and processed using DRAGONS (Data Reduction for Astronomy from Gemini Observatory North and South). 

Based on data obtained from the ESO Science Archive Facility.

% \facility{Swift}

\end{acknowledgments}

%% in the research for the paper. Each keyword is check against the master 
%% list during copy editing. Individual instruments can be provided in 

\vspace{5mm}
\facilities{Gemini, VLT, \textit{Swift}, \textit{CXO}}


\software{DRAGONS \citep{2023RNAAS...7..214L},  
          }



%\appendix

%\section{Appendix information}


\bibliography{NGC300ULX1}{}
\bibliographystyle{aasjournal}



\end{document}